\documentclass{PoS}

\usepackage{verbatim}

\newcommand{\apj}{ApJ}

\newcommand{\prd}{PhRvD}
\newcommand{\aap}{A\&A}

\newcommand{\jgr}{JGR}

\title{%Cosmic rays and their modulation in the inner heliosphere by looking at the gamma-ray Sun}
Solar gamma rays and modulation of cosmic rays in the inner heliosphere}

\ShortTitle{%CR in the inner heliosphere and the gamma-ray Sun}
Solar gamma rays and modulation of cosmic rays}

\author{\speaker{Elena Orlando}\\%\thanks{A footnote may follow.}\\
        Hansen Experimental Physics Laboratory and Kavli Institute for Particle Astrophysics and Cosmology, Stanford University, Stanford, CA 94305, USA\\
        E-mail: \email{eorlando@stanford.edu}}

\author{Nicola Giglietto\\
        Dipartimento Interateneo di Fisica dell'Universita e Politecnico di Bari and INFN Sezione di Bari}
\author{Igor Moskalenko\\
        Hansen Experimental Physics Laboratory and Kavli Institute for Particle Astrophysics and Cosmology, Stanford University, Stanford, CA 94305, USA}
        
\author{Silvia Raino'\\
        Dipartimento Interateneo di Fisica dell'Universita e Politecnico di Bari and INFN Sezione di Bari}

\author{Andrew Strong\\
        Max-Planck Institut f\"ur extraterrestrische Physik, Postfach 1312, D-85741 Garching, Germany}              

\abstract{
The first evidence of the gamma-ray emission from the quiescent Sun was found in the archival EGRET data that was later confirmed by Fermi-LAT observations with high significance. 
This emission is produced by Galactic cosmic rays (CRs) penetrating the inner heliosphere and interacting with the solar atmosphere and optical photons. The solar emission is characterized by two spatially and spectrally distinct components: (i) disk emission due to the CR cascades in the solar atmosphere, and (ii) spatially extended inverse Compton (IC) emission due to the CR electrons scattering off of solar photons. The intensity of both components associated with Galactic CRs anti-correlate with the level of the solar activity being the brightest during solar minimum. \\
In this paper we discuss updates of the models of the IC component of the emission based on CR measurements made at different levels of solar activity, and we make predictions for e-ASTROGAM and AMEGO, proposed low-energy gamma-ray missions.}

\FullConference{35th International Cosmic Ray Conference --- ICRC2017\\
		10--20 July, 2017\\
		Bexco, Busan, Korea}

\begin{document}

\section{Introduction}

%\section{Solar disc and extended diffuse emission}

New materials and technologies that became available to astrophysicists in the last decade enabled CRs and gamma-ray measurements with unmatched precision, which allows for searches of subtle signatures of new phenomena. The launch of PAMELA in 2006, followed by the {\it Fermi} Large Area Telescope ({\it Fermi}-LAT) in 2008, and the AMS--02 in 2011 signify the beginning of a new era in astrophysics. The first ever direct measurements of CRs in the interstellar space became possible due to the high redundancy of {\it Voyager~1} spacecraft launched in 1977 that survived in the harsh space environment for 40 years \cite{Voyager}. These missions contribute a great deal to our understanding of the origins of CRs and the local interstellar medium \cite{Boschini}. Other high-expectations missions are about to release their preliminary data (CALET and DAMPE) or are counting down on the launch pad (ISS-CREAM). Indirect measurements of the CR spectra, outside the influence of the solar modulation, are obtained by looking at the gamma-ray interstellar emission ( e.g. \cite{diffuse1, diffuse2}) and at the interstellar synchrotron emission in radio and microwave frequencies ( e.g. \cite{Strong2011}).

Meanwhile, the direct CR measurements in the close proximity of the Sun, that is just about 1 A.U. away from the Earth, are the most challenging if possible at all. Observations of the gamma-ray emission generated by CR cascades in the solar atmosphere and spatially extended IC emission can be used to deduce the spectra of CR protons and electrons in the inner heliosphere, but require appropriate modeling \cite{Seckel91,Moskalenko, Orlando2006, Orlando2007}.

\section{Theoretical predictions: solar disk and extended emission}

The quiescent Sun as a target for future gamma-ray telescopes was briefly mentioned by Hudson \cite{Hudson} in 1989, but no estimate of the expected gamma-ray flux was attempted at that time. The first calculation of the gamma-ray emission from CR cascades in the solar atmosphere was made by Seckel et al. \cite{Seckel91}, who put forward two models supposedly bracketing the expected flux. The expected flux estimate was high enough to warrant its detection by the EGRET telescope on board of the CGRO mission. Meanwhile, the EGRET team was only able to provide an upper limit on the gamma-ray flux from the solar disk $2 \times 10^{-7}$ cm$^{-2}$ s$^{-1}$ ($>$100 MeV) \cite{Thompson}, while a similar process -- the gamma-ray emission due to the CR cascades in the lunar surface -- was clearly detected. The extended IC emission from scattering of CR electrons off solar photons was overlooked, never mentioned nor searched for by the gamma-ray telescopes.

Only in 2006 it was realized that the IC emission produced in the large heliospheric volume is bright and has a broad distribution on the sky, effectively covering the whole sky \cite{Moskalenko, Orlando2006, Orlando2007}. Even at large elongation angles its brightness is comparable to the brightness of the diffuse Galactic emission. Because of their association with the Galactic CRs, the brightness of both components of the solar gamma-ray emission was predicted to vary over the solar cycle and should anti-correlate with the solar activity. Observations of the IC emission probe the CR electron spectra at different distances throughout the entire inner heliosphere, and allow comprehensive studies of the solar modulation to be performed even in the close proximity of the Sun. Since both components of the solar emission are bright and the Sun is moving across the sky, they have to be taken into account when analyzing the Fermi-LAT data. The appropriate routine has been developed \cite{icrc0957} that became a part of the standard Fermi Science Tools distribution.

\section{The discovery with EGRET and detailed observations with Fermi-LAT}

The first evidence of the gamma-ray emission from the quiescent Sun was found in the archival EGRET data \cite{Orlando2008}. The study was done in the Sun-centered system accounting for the effects of the emission from 3C 279, the Moon, and other gamma-ray sources, which proximity to the ecliptic interferes with the solar emission. Both the disk and the extended IC emission components were found in this analysis. The spectrum and the distribution of the IC component was found to be in agreement with model predictions \cite{Moskalenko, Orlando2006, Orlando2007}, while the flux of the disk component was closer to the ``naive'' model \cite{Seckel91} and exceeding the ``nominal'' one.

The launch of the Fermi gamma-ray telescope in 2008 made observations of the quiet Sun with high statistical significance and on a daily basis a reality.  During the first two years of the Fermi mission the solar activity was extremely low, resulting in a high heliospheric flux of Galactic CRs. Therefore, the CR-induced quiescent gamma-ray emission from the Sun was expected to be near its maximum. The results of the analysis of observations of the solar emission during the first 18 months of the Fermi LAT mission were reported in \cite{Abdo2011}. The data analysis was done in Sun-centered system, while the background was estimated using the so-called ``fake-Sun'' method that uses an imaginary source that is trailing the Sun along the ecliptic. We used averaging of the backgrounds derived independently from 4 fake-Sun sources displaced from each other and from the Sun itself by $40^\circ$. The 4 backgrounds were anyway very similar and were used to validate the method. The observed disk spectrum was fitted by a single power law with a spectral index of $2.11\pm0.73$. The observed integral flux from the solar disk was found to be $(4.6 +/- 0.2$ [statistical error] $+1.0/-0.8 $[systematic error]$)\times10^{-7}$ cm$^{-2}$ s$^{-1}$ ($>$100 MeV), which is $\sim$7 times higher than predicted by the ``nominal'' model \cite{Seckel91}, and consistent with results of the previous analysis of the EGRET data \cite{Orlando2008}. 
One of the reasons for such a discrepancy of the solar disc emission with the predicted flux \cite{Seckel91} was unusually deep solar minimum during the reported observations. However, this alone cannot account for such a large factor, see a comparison with the EGRET results in \cite{Orlando2008}. Another possibility for an estimated ``nominal'' flux to be so low compared to the Fermi LAT observations is that the CR cascades developing in the solar atmosphere could be bent backwards toward the surface by the much stronger than expected magnetic field in the tubes below the chromosphere. A miscalculation of the ambient flux of CR protons could also be a possibility. Accurate measurements of the disk spectrum by the Fermi-LAT thus warrant a new evaluation of the CR cascade development in the solar atmosphere.
In contrast, the observed integral flux of the extended IC emission from a region of 20$^\circ$ radius centered at the Sun is $(6.8 +/-0.7$ [stat.] $+0.5/-0.4$ [syst.]$) \times10^{-7}$ cm$^{-2}$ s$^{-1}$ (>100 MeV), in a good agreement with expectations. The observed angular profile of the IC emission is also in a good agreement with theoretical predictions. Although the highest energy point 3-10 GeV shows some excess relative to the model predictions, this is difficult to explain from the model viewpoint since the effect of the solar modulation is decreasing at high energies thus making the model more accurate. Meanwhile, the statistical error bars are still too high that does not allow to discriminate between models predicting different behavior of the electron spectra at different heliospheric distances. A new analysis with higher statistics, finer angular and energy grids is highly desirable. 

A recent analysis of 6 years of Fermi-LAT observations \cite{Ng} reported variations of the flux of the disk component that anti-correlates with the solar activity. However, the IC component that is more difficult to analyze was not the subject of that study. 

Preliminary results of the on-going analysis of the two components of the solar emission have been presented in \cite{Raino}. The gamma-ray flux from the Sun was evaluated using Pass 8 data collected during the first 7.5 years of the Fermi-LAT operation. Significantly larger photon statistics and improved processing performance with respect to the previous analyses is allowing us to explore both components of the emission in greater details and discriminate between current models of the IC emission.

\section{Updated models consistent with the latest CR data}

The first models of the IC emission \cite{Moskalenko, Orlando2006, Orlando2007} employed the CR all-electron data available at that time. However, new precise CR electron measurements by AMS-02 \cite{AMS_ele} has become available since that time which show a noticeably softer all-electron spectrum that also has larger flux at low energies. Therefore, the models have to be updated according to these new measurements. Hence, the new models incorporate the local interstellar all-electron spectrum derived through the demodulation of the AMS-02 data. % consistent with Voyager 1 data at low energies. 
%Models were generated using the StellarICs code\footnote{Available from https://gitlab.mpcdf.mpg.de/aws/stellarics} \cite{Orlando2013}, which has now been updated with this formulation. 
The corresponding numerical implementation of the models, the StellarICs code\footnote{Available from https://gitlab.mpcdf.mpg.de/aws/stellarics} \cite{Orlando2013}, has now been updated. 

Figure~1 shows the all-electron spectrum modulated with force-free modulation potentials of 400 MV and 600 MV compared to the data. Figure 2 shows the modeled IC flux integrated over the circular areas with different elongation radii. The predictions of the IC flux are based on the all-electron spectrum tuned to AMS-02 data as described above.

\begin{figure}
\center
\includegraphics[width=0.7\textwidth]{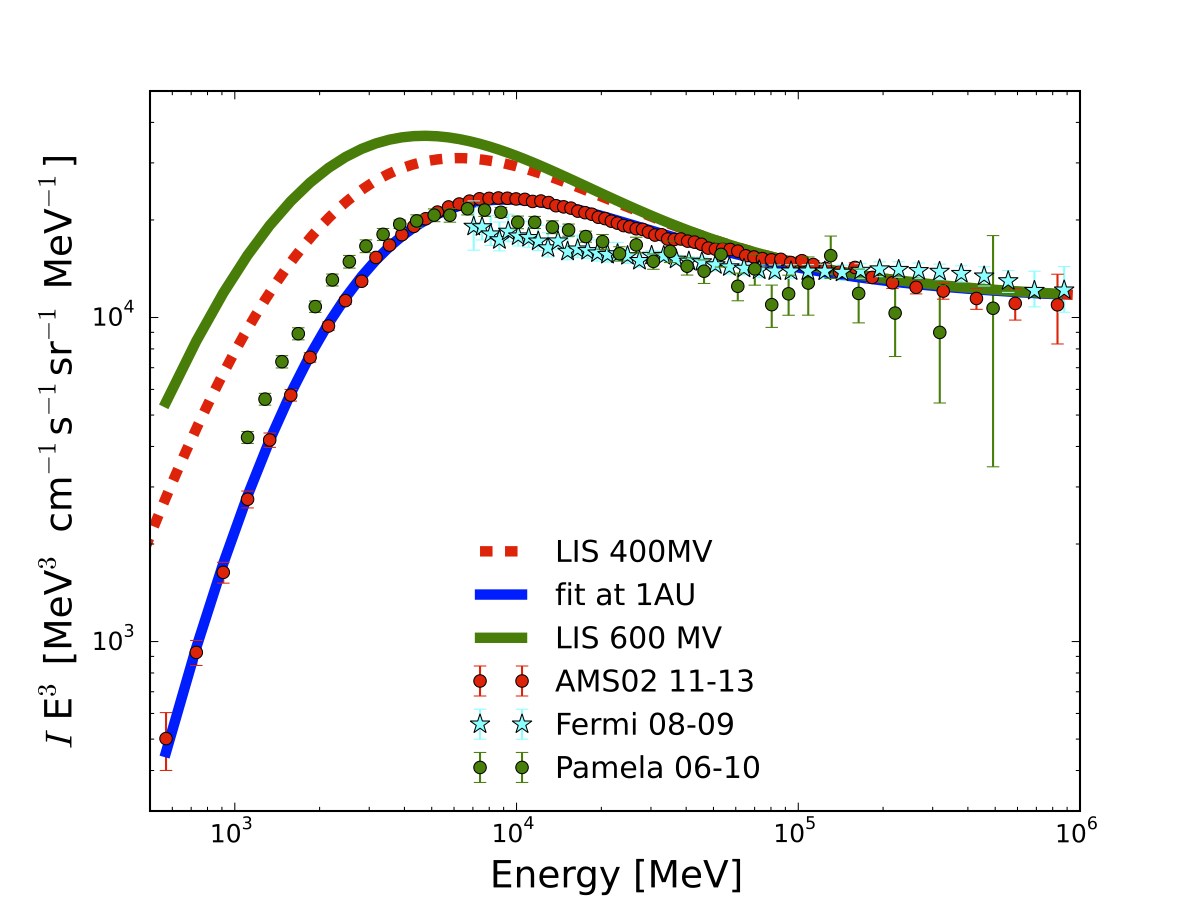}
\caption{All-electron spectra for modulation potential values of 400 MV and 600 MV compared with AMS-02 \cite{AMS_ele}, Fermi \cite{fermi_ele}, and Pamela \cite{Pamela_ele} data. The all-electron spectrum is tuned to the AMS-02 measurements.}
\end{figure}  

\begin{figure}
\center
\includegraphics[width=0.7\textwidth]{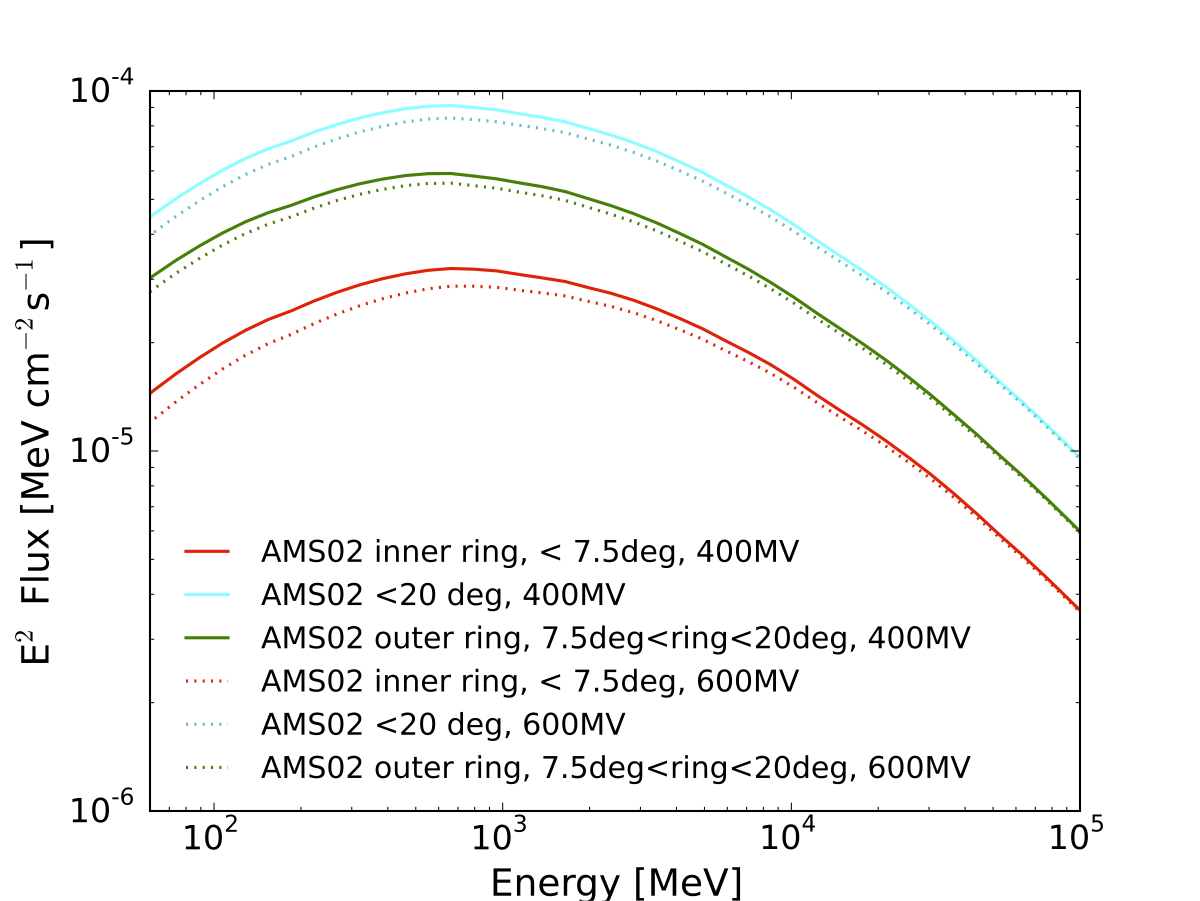}
\caption{Calculated IC flux integrated over areas with different elongation radii for the all-electron spectra shown in Figure 1.}
\end{figure}  

\begin{figure}
\center
\includegraphics[width=0.7\textwidth]{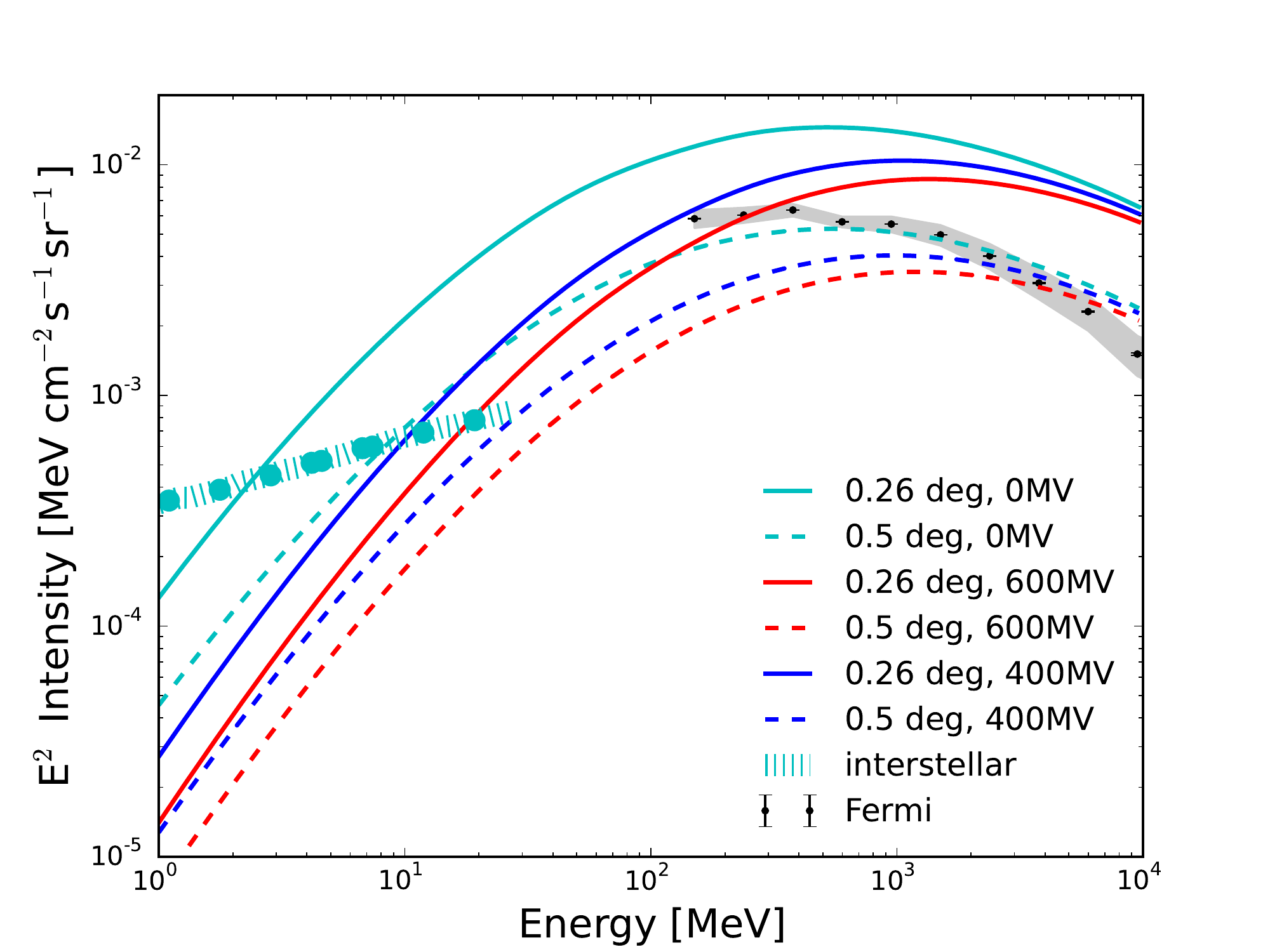}
\caption{Model predictions of the intensity of the solar IC emission in the energy range of interest for proposed e-ASTROGAM and AMEGO missions. Calculated IC flux integrated over areas with different elongation radii for the all-electron spectra shown in Figure 1. Shown also the Fermi LAT data \cite{diffuse1} and predictions of the diffuse Galactic emission at intermediate latitudes.}
\label{fig_eAstrogam}
\end{figure}

\section{Energy range of interest for e-ASTROGAM and AMEGO missions}

The solar IC emission is also contributing to lower energies, which will be observed by the proposed MeV-GeV energy missions, such as e-ASTROGAM \cite{eAstrogam} and AMEGO\footnote{https://asd.gsfc.nasa.gov/amego/}. In more detail, e-ASTROGAM is designed to detect photons from 0.3 MeV to 3 GeV, while AMEGO range is 0.2 MeV to 10 GeV. Therefore, our predictions are extended down to 1 MeV for different models. Figure \ref{fig_eAstrogam} shows the predictions for the range of interest for e-ASTROGAM and AMEGO missions. Given the broad distribution of the IC emission on the sky, the contribution of the solar IC emission to the diffuse background cannot be neglected. On the other hand, such measurements will enable us to study the low-energy part of the all-electron spectrum in close proximity of the Sun.

\section*{Acknowledgments}
Elena  Orlando acknowledges support from NASA grant No.~NNX16AF27G. Igor Moskalenko acknowledges partial support from NASA through grants Nos.~NNX13AC47G and NNX17AB48G.


\begin{thebibliography}{99}
\bibitem{diffuse1} Abdo, A.~A., Ackermann, M., Ajello, M., et al.\ 2009, Physical Review Letters, 103, 251101
\bibitem{Abdo2011} Abdo, A.~A., Ackermann,  M., Ajello, M., et al.\ 2011, ApJ, 734, 116 
\bibitem{diffuse2} Ackermann, M., Ajello, M., Atwood, W.~B., et al.\ 2012, \apj, 750, 3 
%\bibitem{diffuse2} Ackermann, M., 
%Ajello, M., Atwood, W.~B., et al.\ 2012, \apj, 750, 3 
%\bibitem{Abdo_cygnus} Ackermann, M., Ajello, M., Allafort, A., et al.\ 2011, Science, 334, 1103 
\bibitem{fermi_ele} Ackermann, M., Ajello, M., Atwood, W.~B., et al.\ 2010, \prd, 82, 092004 
\bibitem{Pamela_ele} Adriani, O., Barbarino, G.~C., Bazilevskaya, G.~A., et al.\ 2015, \apj, 810, 142
%\bibitem[Aguilar et al.(2013)]{AMS-02} Aguilar, M., Alberti, G., Alpat, B., et al.\ 2013, Physical Review Letters, 110, 141102 
\bibitem{AMS_ele} Aguilar, M., Aisa, D., Alvino, A., et al.\ 2014, Physical Review Letters, 113, 121102

\bibitem{Boschini} Boschini, M.~J., Della Torre, S., Gervasi, M., et al.\ 2017, \apj, 840, 115
\bibitem{Voyager} Cummings, A.~C., Stone, E.~C., Heikkila, B.~C., et al.\ 2016, \apj, 831, 18 
\bibitem{eAstrogam} De Angelis, A., Tatischeff, V., Tavani, M., et al.\ 2017, Experimental Astronomy, 
\bibitem{Hudson} Hudson,  H.  S.,   1989,  Proc  Gamma  Ray  Observatory  Workshop  (Greenbelt: Goddard Space Flight Center), 4-351
\bibitem{icrc0957} Johanesson, G., \& Orlando, E.,\ 2013, Proc. 33rd International Cosmic Ray Conference, p.0957 (arXiv:1307.0197)
%\bibitem{Moska2000} Moskalenko, I.~V., \& Strong, A.~W.\ 2000, ApJ, 528, 357 
\bibitem{Moskalenko} Moskalenko, I.~V., Porter, T.~A., \& Digel, S.~W.\ 2006, ApJL, 652, L65 
\bibitem{Ng} Ng, K.~C.~Y., Beacom, J.~F., Peter, A.~H.~G., \& Rott, C.\ 2016, \prd, 94, 023004 

\bibitem{Orlando2006} Orlando, E., \& Strong, A.~W.\ 2006, arXiv:astro-ph/0607563
\bibitem{Orlando2007} Orlando, E., \& Strong, A.~W.\ 2007, Ap\&SS, 309, 359 
\bibitem{Orlando2008} Orlando, E., \& Strong, A.~W.\ 2008, A\&A, 480, 847 
%\bibitem{Orlando2009} Orlando, E., \& Strong, A.~W.\ 2008, International Cosmic Ray Conference, 2, 505 
\bibitem{Orlando2013} Orlando, E., \& Strong, A.\ 2013, Nuclear Physics B Proceedings Supplements, 239, 266 
%\bibitem{Orlando2013mnras} Orlando, E., \& Strong, A.\ 2013, \mnras, 436, 2127
\bibitem{Raino} Rain{\'o}, S., Giglietto, N., Moskalenko, I., Orlando, E., \& Strong, A.~W.\ 2017, European Physical Journal Web of Conferences, 136, 03007 
\bibitem{Seckel91} Seckel, D., Stanev, T., \& Gaisser, T.~K.\ 1991, \apj, 382, 652 
\bibitem{Strong2011} Strong, A.~W., Orlando, E., \& Jaffe, T.~R.\ 2011, \aap, 534, A54 
\bibitem{Thompson} Thompson, D.~J., Bertsch, D.~L., Morris, D.~J., \& Mukherjee, R.\ 1997, \jgr, 102, 14735 

\end{thebibliography}
\end{document}